\documentclass[a4paper]{article}

\usepackage{sctr}
\usepackage{times}
\usepackage{url}

\usepackage{amssymb}

\usepackage{graphics}
\usepackage{color}
\newcommand{\comment}[1]{}
\definecolor{Orange}{rgb}{1,0.5,0}

\title{Optimizing a Personalized Multigram Cellphone Keypad}

\author{Joonseok Lee\\
	R I (Bob) McKay}
\email{(joonseok2010,rimsnucse)@gmail.com}

\date{2014/11/29}

\report{TRSNUSC:2014:001}

\begin{document}

\maketitle			

\pagestyle{empty}

\clearpage

\mbox{}

\clearpage

\pagenumbering{roman}

\abstract
Current layouts for alphabetic input on mobile phone
keypads are very inefficient. We propose a genetic algorithm
(GA) to find a suitable keypad layout for each user, based on their
personal text history. It incorporates codes for
frequent  multigrams, which may be directly input.
This greatly reduces the average number of strokes required
for typing.

We optimize for two-handed use, the left thumb covering the
leftmost rows and vice versa. The GA minimizes the number of
strokes, consecutive use of the same key, and consecutive use of the
same hand. Using these criteria, the algorithm re-arranges the 26
alphabetic characters, plus 14 additional multigrams, on the 10-key
pad. We demonstrate that this arrangement can generate a more
effective layout, especially for SMS-style
messages. Substantial savings are verified by both computational
analysis and human evaluation.

\cleardoublepage

\pagestyle{plain}

\pagenumbering{arabic}

\setcounter{page}{1}

%#################################### Begin Main Content ##############################################

\section{INTRODUCTION}
\label{sec:intro} Current cell-phone keypad layouts
are of two kinds: unambiguous and ambiguous.
Unambiguous layouts have enough keys to uniquely map the intended
alphabet to the keys -- the QWERTY keyboard is an example. This
is straightforward to learn, but difficult to use because keys are
too small for finger use due to the lack of
space on mobile phones.

Thus ambiguous layouts, with fewer keys than
alphabetic characters, are widely used. Two mechanisms are
used to resolve the ambiguity as to which
letter is intended: multi-tap and contextual disambiguation.
Multi-tap uses repetition: letters on the same key are
distinguished by number of strokes. The most widely-used
12-key layout falls into this
category. Although easy to learn due to the natural ordering,
it is inefficient to use -- most obviously because of the high
average number of key strokes. In addition, it performs poorly for
single-thumb text entry because of the high average distance the
thumbs must move~\cite{moradi:genetic}. But most important, a 12-key
pad only permits us to type one character per key
press~\cite{gong:alphabetical,lesher:ambiguous,moradi:genetic,nesbat:fastfull}.
Since English (like most languages) has many common digrams and
trigrams, a pure unigram encoding misses important
opportunities for efficiency.

The other alternative is disambiguation algorithms. In these, the
user types only a single stroke per letter, regardless of the
position of the intended letter on the key. The disambiguation algorithm
guesses the intended word using statistical characteristics
of the language. For example, when the user types key sequence '63',
the algorithm guesses the intended word as 'of', based on the
frequency of English words. If the user does not intend 'of', the
algorithm shows the next most probable word, 'me'. However predictive
methods are not a panacea. Nesbat~\cite{nesbat:fastfull} argues that
disambiguation methods create unreasonable cognitive loads: the user
has to continually check whether the intended word has been
recognized, and edit it if not. The problem is even worse when the
intended words are not in the dictionary. The user must change to
direct input mode and re-type. This case, however, is surprisingly
frequent, especially in SMS. Because of the restricted message
length, users often contract words to abbreviations: 'btw' for 'by
the way', for example. They also commonly use proper names for
people or places. Nesbat also notes that users frequently mix
languages, as in 'Hello Amigo'; thus predictive methods
based on standard dictionaries perform unexpectedly poorly.

We aim to overcome these limitations 
by supporting personalization and multigrams.
Optimization based on a personal SMS archive has many merits. Most
important, a personalized layout is efficient. Ideas for remapping
keypad layouts to date have assumed a standardized layout, and used a
standard frequency table. However, letter and bigram frequency
differs user by user, culture by culture, despite use of the same
language. To be specific, American, British, and Australian usages
differ in word frequency, spelling, and SMS practice, although all
use English. Reflecting and taking advantage of these differences
can increase speed. For instance, 'q' is more frequently used in
Canada, because of its French culture. A personally-optimized layout
may allow this generally-rare lettter to be on a better position.
Personalization allows us to take advantage of each user's frequent
words or phrases -- even those containing foreign languages (e.g.
amigo), proper nouns (e.g. SNU, LG), abbreviation (e.g. btw, 4u),
slang (e.g. wusup), and even emoticons (e.g. :), \^\--\^ \-\ ). A
personalized character input system has further potential. It may
support assistive technology by reflecting a specific user's
difficulty in typing. A personalized layout can also work as a
security tool by recognizing the owner's usual pattern of use.

Another key idea in this paper is multigrams. Frequently-used
bigrams or trigrams are used in the layout, inputting more than one
letter at once. Note that the list of multigrams is also
personalized: frequently-used ones in the individual archive
are selected. In a sense, selecting multigrams can be seen as a
prediction algorithm (multigram, instead of word). We believe that
selecting personally-frequent multigrams improves prediction
accuracy compared to a standard word prediction system. In the latter,
the user must check whether the intended word is
chosen and edit it if not. In a multigram layout, this
risk is low once the user sufficiently memorizes his
own layout.

We constrain this work to English 12-key layout, although the method
can be readily extended to other single-byte character systems like
Spanish. Only 10 keys are used for typing, with two keys reserved
for special purposes, such as input of special characters and mode
change. Upper and lower case in the user archive is ignored because case
is distinguished generally by mode, not by key. We also assume
availability of memory to record the user's keystroke history.

In the rest of the paper, we provide the historical context relating
to keyboard layout optimization and personalization first, followed
by a detailed description of our methods. Then, we detail the
evaluation design and results, using a computational analysis,
following this with a human evaluation. Lastly, we conclude with a
summary of the contribution, limitation and our planned future work.

\section{BACKGROUND}
\label{sec:back}
\subsection{Optimizing Mobile Keypads}

Many researchers have tried to optimize mobile keyboards. We can
classify their work firstly by the kind of physical keypad. For
ambiguous keypads, MacKenzie and Zhang~\cite{mackenzie:design}
designed a soft keyboard OPTI, redesigning the keyboard shape and
key distribution. Zhai et el.~\cite{zhai:metropolis} also designed a
virtual keyboard for touch screens, optimizing with the Metropolis
algorithm. They measured the efficiency of several existing virtual
keypads including QWERTY. Despite this work, unambiguous keypads are
little used in mobile phones, probably due to space limitations
for keys.

By contrast, multi-tap and disambiguation are widely used in
cell-phones. For predictive systems, Lesher et
al.~\cite{lesher:ambiguous} optimized the
character-level disambiguation rate for a given text, using the
n-opt algorithm (well-known from Traveling Salesman
Problems -- TSP). The keystrokes per character (KSPC) dropped to 1.09. 
Gong and Tarasewich
~\cite{gong:alphabetical} added another
constraint: assigning characters in alphabetical
order. They observed KSPCs ranging from 1.05 (written article) to
1.25 (text messages on mobile phone), for 10 keys. This resulted
from a disambiguation rate of up to 98.13\% for written text and
95.13\% for SMS. 

Genetic Algorithms (GA) have often been used to optimize layouts. How and
Kan~\cite{how:predictive} remapped key layout with a simple GA. They
also tried to improve accuracy of word prediction by investigating
previous words in addition to the current key sequence.

GAs have also been used for remapping keys for multi-tap
layouts. Moradi and Nickabadi~\cite{moradi:genetic} use a GA to
optimize, based on a frequency sample. For a given sample, their GA
minimized average strokes, delay due to consecutive use of the same
key, and moving distance to type the given sample, with an
assumption that only one thumb is used for typing. In their metric,
typing cost dropped from 2.90 (current layout) to 2.25 (optimized).
They stated that the theoretical minimum of the cost function is
1.70. We note that these cost values depend heavily on
characteristics of the text, including the variety of words, the
kinds of words, and the occurrence of special characters. We use
their work as a basis for comparison with our new, multigram
approach.

\subsection{Personalized Keyboards}
All these methods 
implicitly assume that keyboard design should be standardized.
Nguyen et al.~\cite{mckay:chording} argue that universal keyboards
are unnecessary due to the rise of standard interfaces, such as USB,
W-USB, and Bluetooth. For both computer and mobile phone input,
personalized key layouts can be readily downloaded into such
devices. They propose a GA to generate a user-optimal key layout for
10-finger pure-chording devices.

\subsection{Genetic Algorithms}
GAs are powerful methods for tough search spaces, performing
well on most problems, and being relatively easy to use. 
We refer readers to a standard text such as De
Jong's~\cite{EvolutionaryComputation} for a detailed background.
They may be divided into two classes, generational (at each
stage, the parent population generates a new, separate population
constituting the next generation), and steady-state
(individual parents produce children, which then compete with parents
to replace them).

We use GAs in two places: choosing multigrams for
 the layout, and choosing the arrangement of
keys. 
We use steady-state GAs, in which one (mutation) or two
(crossover) parents are chosen randomly from the current
population. The relevant operator is applied, producing a child, which
 competes deterministically (truncation selection) with
the parent(s), the least fit being eliminated from the population.

\section{MOTIVATION}
\label{sec:mot}
\subsection{Criteria for Keypad Optimization}
We assume that both thumbs are used for typing.
This  is the fastest mode for SMS input from a
keypad, and thus automatically focuses on those
users most interested in speed, and most likely to
adopt a new input mode.
Thus unlike Moradi and Nickabadi's~\cite{moradi:genetic},
our fitness function emphasizes the number of
key strokes rather than distance (because
distances for two-thumb operation are generally small).

In this context, three criteria determine the efficiency of a
layout. Most important, fewer key strokes per character improve
input speed. We need between one and three strokes to type a
character in current systems. For example, we use three strokes to
type "F" but only one for "G". If frequent characters use the
first-stroke position, we would reduce the overall number of
strokes. But frequent characters vary from person to person, so
reflecting the personal history is important in minimizing the
number of strokes.

Second,
if we want to type another letter using
the same key, we must include a delay
or type an additional stroke
to move the cursor -- either slows typing.
Thus reducing the probability of consecutive use of
the same key, based on the personal
archive, will improve typing efficiency.
Moradi also uses this criterion,
with a slightly different weight factor.

Finally, for two-thumb use, alternating thumbs eliminates
thumb movement from consideration: the alternate thumb can
move into position while the first makes a stroke. Thus our
penalty should only apply when the same thumb must be used for
consecutive letters.

We need to combine these three criteria, weighting them to
reflect their impact on typing speed.
A more formal description is given below.
We also incorporate multigrams in the layout. While
these improve typing efficiency, they do not
introduce any further complications to the fitness function.

\subsection{Formal Description}

Using these criteria, we define a three-part fitness function:
\begin{equation}
f_1 = \frac {\sum_{l} st(l)} {C}
\end{equation}
where  $st(l)$ is the number of strokes required to type character
$l$, and $C$ is the total number of characters in the corpus.
\begin{equation}
\label{eq:sk}
f_2 = \frac {\sum_{l} sk(l)} {C}
\end{equation}
where $sk(l)$ is 1 if consecutive characters use the same key, else
0. In the text "What is this?", if 't' and 'h' are located on one
key and 'i' and 's' on another, the numerator of
equation~\ref{eq:sk} would be 3, because t-h, i-s, and the second
i-s share the same key. Thus equation~\ref{eq:sk} gives a penalty of
0.3.
\begin{equation}
f_3 = \frac {\sum_{l} sh(l)} {C}
\end{equation}
where $sh(l)$ is 1 if consecutive characters use the same hand, 0
otherwise. Thus the penalty for LLRLRLR is 1, and 3 for RLLLRRL.
However, there is a complication.
We assume that keys on the left
column
use the left hand, and vice versa. But the middle column may be
typed with either hand -- whichever was not used last. We can define
the parity of a sequence of middle column characters as follows:
when we have two non-center characters using the same hand,
separated by an even number of middle-row characters, or else
opposite hands separated by an odd number, we say they have even
parity, otherwise odd. When the parity is even (i.e. the character
stream cannot be typed with alternating hands), the penalty is 1,
otherwise it is 0.
Of course, this
assumes foresight of arbitrary length on the part of the typist. In general,
long sequences of central characters will be rare, so the
effect of limited foresight will be small.

The overall adaptive fitness
function is defined as
\begin{equation}
F_A = \alpha f_1 + \beta f_2 + \gamma f_3
\end{equation}
Normalizing by setting $\alpha = 1$,
we can rationally choose relative values for the other terms.
$f_2$ penalizes repeated use of the same key. Speed-conscious
typists will use the cursor key rather than a timeout, so the cost is one stroke; but
the cursor key is located inconveniently (compared to other keys),
so we set $\beta=1.5$.
For $f_3$, the effect of same-hand use is
relatively minor, so we use $\gamma=0.25$.

In comparisons with Moradi and Nickabadi, we use their distance term
\begin{equation} f_4 = \frac
{\sum_{l} d(l, prev(l))} {C}
\end{equation}
with $d(a, b)$ defined as $\sqrt{(r_a - r_b)^2 + (c_a -
c_b)^2}$, with $r$ and $c$ denoting row and column. Their
cost function is then:
\begin{equation}
F_M = \alpha_M f_1 + \beta_M f_2 + \gamma_M f_4
\end{equation}
where $\alpha_M = 0.7$, $\beta_M = 3$ and $\gamma_M = 1$ as
in~\cite{moradi:genetic}.

\subsection{Multigrams}
To further improve typing speed (and reduce $f_1$), we introduce
multigrams in our coding, and assign key positions for
frequently-used multigrams. Thus one key stroke may create more than
one character. In this paper, we allow at most four strokes for each
key. Using ten keys, this gives 40 available slots. Because there
are 26 letters in the English-Roman alphabet, 14 slots are available
for multigrams. Thus we first choose the best combination of the 26
letters and 14 multigrams to minimize the overall cost function.

It might seem that we could simply select the 14 most frequent
multigrams. However two key issues preclude this. First, the most
frequent bigrams may overlap the most frequent trigrams. For many
corpora, "er" and "ert" will be frequent. However if we include
"ert" in our coding, other occurrences of "er" may be too rare to
justify a separate coding. Second, multigrams will change the
unigram frequencies of their constituent letters. For example, if
"of" occurs 30 times, "o" occurs 100 times, and "f" occurs 50 times,
then a multigram for "of" covers 30\% of "o" and 60\% of "f". This
reduction in the frequencies of "o" and "f" may result in their
assignment to low key positions (requiring more key strokes),
freeing up their otherwise-higher positions for other keys which are
now more frequent.

Instead of selecting the 14 most-frequent multigrams, we assume the
relevant ones will be among the 50 commonest bigrams and trigrams of
the archive. To select the 14 multigrams in the coding, we
pre-process with an initial GA. Our variant of GA uses an eager
initialization: the 14 multigrams are selected by roulette sampling
based on their frequency of occurrence (i.e. the probability of
selection is proportional to the frequency), where a typical GA
would use uniform sampling. It also uses eager operators. The
crossover operator preserves all multigrams that occur in both
parents, then selects other multigrams from either parent in order
of frequency (a more typical crossover would randomly select the
multigrams from either parent). The mutation operator randomly, but
at a low rate (1\%) replaces the 14 multigrams with others from
outside the top 50 (in order to improve the solution when our
initial assumption was wrong). For the fitness function, we use the
sum of the changes in rank of the unigrams, since this tells us how
often the multigrams will be located on high-priority key positions,
and hence serves as a proxy for the key strokes they will save. We
use deterministic selection between a child and its parent(s). We run
the GA for 1000 repetitions, with a population of 50, which seems to
be sufficient to give convergence to an optimum.

Once we have determined the 14 multigrams, we treat
them exactly like ordinary letters, letting them compete for fewer
strokes and better positions. If the frequency of a multigram
 is low, it may be allocated a bad position such as 4 strokes.
 If this results in more strokes than direct input would require,
 the multigram is deprecated.

\section{THE CORE GENETIC ALGORITHM}
\label{sec:meth} The search for the best arrangement of the alphabet
(including the chosen multigrams) on the keypad is not easy. This is
a rough search space with many local minima, and with
difficult-to-satisfy constraints. Straightforward optimization
algorithms will fail, repeatedly generating infeasible solutions. We
use a steady-state GA to find a good arrangement.

A GA can be described by specifying five components -- the genotype
representation, the initialization process, and the crossover,
mutation and selection operators -- together with a number of
parameters detailing the exact set-up.

\subsection{Chromosome Expression}
Our chromosome  has the following structure:
\begin{verbatim}
public class Solution {
    private Gene[40] gene;
    private int size;
    private int id;
} public class Gene {
    private int row;
    private int column;
    private int stroke;
}
\end{verbatim}
Thus there are 40 genes, one for each letter. Each gene gives the
assigned location for the corresponding letter (row, column,
strokes). Corresponding to the 4 rows and 3 columns of the keypad,
row has the range 1 to 4, and column 1 to 3. The strokes value
specifies the number of strokes to type the character, also 1 to 4.
Table \ref{table:chromosome} shows a typical chromosome.

\begin{table}
\begin{centering}
\begin{tabular}{|c|r|r|r|r|r|r|r|r|r|r|r|r|r|}
\hline
  & 0 & 1 & 2 & 3 & 4 & 5 & 6 & 7 & 8 & 9 & .. & 39\\
\hline
R & 2 & 2 & 1 & 3 & 3 & 1 & 2 & 2 & 1 & 4 & .. & 4\\
C & 3 & 2 & 2 & 3 & 1 & 2 & 2 & 1 & 1 & 2 & .. & 2\\
S & 4 & 1 & 1 & 2 & 2 & 2 & 2 & 1 & 1 & 1 & .. & 3\\
\hline
\end{tabular}
\par\end{centering}

\caption{Example Chromosome} \label{table:chromosome}
\end{table}

There are two restrictions. Because we use only 10 of the 12 keys
excluding key '$*$' and '\#', two locations (4,1) and (4,3) should
not occur. So genes such as (4, 1, 1) and (4, 3, 3) are forbidden.
Also, each gene represents a unique location and number of strokes,
so no duplicates are allowed: every gene should be unique. Crossover
and mutation must maintain these properties.

\subsection{Initialization}
Each individual in the initial population is a permutation of the
letters on the keypad, the permutations being sampled uniformly
randomly.

\subsection{Crossover}
In crossover, we aim to create children lying semantically
\textit{between} parents. We define \textit{between} formally as: a
gene $g_1$ is \textit{between} genes $g_2$ and $g_3$ if each
component (row, column, strokes) lies in the closed interval defined
by the corresponding values in $g_2$ and $g_3$. A chromosome is
\textit{between} two others if all corresponding genes satisfy this
property. In the early stages of evolution, each element may have a
large range, but we expect it to converge to a specific value
because of the influence of fitness on selection.

To help construct children lying (if possible) \textit{between}
parents, we use two data structures: a 3-dimensional array [4, 3, 4]
of lists \textit{candidate}, expressing the possible keypad
structures; and a 1-dimensional array of 40 integers
\textit{possibleCount}, containing the number of \textit{between}
positions where the corresponding gene can be placed. For each gene
we determine the possible locations that lie \textit{between} the
corresponding genes of the parent. For example, if the parents for
gene $x$ are (1, 2, 3) and (3, 2, 4), the candidate locations are
(1, 2, 3), (1, 2, 4), (2, 2, 3), (2, 2, 4), (3, 2, 3), and (3, 2,
4). We insert $x$ in every corresponding position in the list array
\textit{candidate}, while gene $x$ in \textit{possibleCount} gets
the value 6 (for the six possible locations). This process is
repeated for every gene. Thus \textit{candidate} array contains
every possible gene that can be placed at each location, while
\textit{possibleCount} contains the number of possible locations for
each gene.

Then the arranging algorithm first searches for genes with
\textit{possibleCount} 1. This occurs when both parents have
identical (row, column, stroke) values. Thus the child has only one
possible position for this gene; it is automatically located there.
Thus no other gene can occupy the same position: their
\textit{possibleCount} is reduced by 1.

When all remaining genes have more than one choice, the algorithm
moves on to a gene with only two choices (we call this value, 2, the
\textit{level}). Of its two candidate positions, we choose the one
with fewest remaining possible genes. Thus if one position has 6
possibilities and the other 2, we choose the latter -- again,
eliminating this choice from other genes. The aim is to leave open
as many choices as possible for the remaining genes. Note that this
process may move some other gene to \textit{level} 1 (by eliminating
one of its two choices). Thus at each stage, the algorithm should
re-start from \textit{level} 1, and always choose a gene from the
lowest available \textit{level}. Since this process always increases
\textit{level}s, and since the \textit{level}s cannot exceed the
number of available places, it must terminate in at most $40 + 39 +
38 + ... + 1 = 820$ iterations.

There is one exception. In rare cases, we may not be able to satisfy
the \textit{between} condition - the \textit{possibleCount} may drop
to zero. One option, in this case, is to backtrack and try
alternatives. However this is expensive. Since GAs need a source of
randomness anyway (and indeed, mutation is included just for this
purpose), we adopt a simple solution: when a gene cannot be located
\textit{between} its parents, we allocate it randomly among the
vacant positions, and adjust the data structures correspondingly. In
fact, this is rarely used (less than one case in 100).

\subsection{Mutation}
We used a number of mutation operators: swapping two columns (24 to
28), swapping two rows (24), swapping two keys (8), reorganizing
strokes in a key(4), and exchanging a pair of letters(2). (The
figure in parenthesis means corresponding number of genes affected
by the mutation.) Thus the first two may be viewed as relatively
global mutations, exploring the search space, while the last three
are local mutations, searching the space finely.

\subsection{Selection}
For crossover, selection works by randomly selecting two parents,
creating the child, then holding a tournament between the parents
and the child, with the loser dropping out of the population.
Parents for mutation are chosen randomly, with the child always
replacing the parent.

\section{COMPUTATIONAL ANALYSIS}
\label{sec:comp}

To compare the typing cost of Personalized Multigram (PM)
and ABC layouts, we used a computational analysis followed by human
evaluation. Here, we measure the theoretical improvement
by comparing fitness values.

\subsection{Experimental Design}

\begin{table}[htbp]
\begin{centering}
\begin{tabular}{|c|c|c|l|}
\hline Type & Words & Letters & Description  \tabularnewline \hline
\hline SMS & 554 & 2766 & Recent SMS messages by  \tabularnewline
 & & & the first author
 \tabularnewline
FCB & 725 & 3358 & Instant messages gathered\tabularnewline
 & & & from Facebook\tabularnewline
ART & 378 & 2340 & An article relating to.\tabularnewline
 &  & & a social issue~\cite{time:trip}
\tabularnewline \hline
\end{tabular}
\par\end{centering}
\caption{Profile of Archives used in Experiments
 \protect\\
{\footnotesize The number of characters include empty space.}}
\label{table:archive}
\end{table}

To test effectiveness, we used three kinds of text
(Table~\ref{table:archive}). Each archive came from one author,
reflecting our aim of matching personal use. The first uses recent
Short Message Service (SMS) messages of the first author. The second
uses Facebook (FCB) postings, based on an assumption that language
usage may be similar in SMS and social networks -- in terms of
topics (personal issues dominate), frequently occurring
abbreviations and emoticons. The last is an article (ART) from TIME
magazine~\cite{time:trip}, which should reflect general usage, but
also include characteristics of the author in word choice, voice,
and use of pronouns; but the article does not contain the
abbreviations and emoticons typical of SMS. All characters are
either one of the  26 English letters, or special characters
normally found on phone keypads.

\subsubsection{Parameter Settings} GAs are stochastic;
runs  may give different results. In comparisons, we need
to run multiple trials,  assessing differences
statistically; we used 50 trials for each treatment. It is also
essential to use fair comparisons. Results can be improved by
extra computation so it is important to
compare equal effort. This is usually specified
as the number of potential solutions evaluated (in our case, keyboard
configurations). Parameter Settings
are shown in Table~\ref{table:params}.

\begin{table}
\begin{centering}
\begin{tabular}{|l|r|}
\hline Parameter & Value \\ \hline \hline
Number of Trials & 50\\
Population size & 50\\
Number of Evaluations & 50,050\\
Mutation rate (each of 5 types) & 0.01\\
Crossover rate & 1\\
Algorithm type & steady-state\\
Selection & Parent-child\\
& tournament\\
Elite Size & 1\\ \hline
\end{tabular}
\par\end{centering} \caption{Experimental Parameters}
\label{table:params}
\end{table}

\subsubsection{System Environment} We conducted these experiments on an
Intel Core 2 @ 2.13GHz machine with 2GB RAM. MS Windows XP
Professional SP 3 was used for OS, and our analysis program ran under Java
2 Runtime Environment, Standard Edition 1.4.1.

\subsection {Result and Analysis} \label{sec:res}

\subsubsection {Comparison with Moradi and Nickabadi}

\begin{table}[htbp]
\begin{centering}
{\footnotesize
\begin{tabular}{|c|l|c|c|c|c|c|}
\hline Text & Multi- & ABC  &
\multicolumn{2}{c|}{Random Keypad} & \multicolumn{2}{c|}{Optimized
Keypad}\tabularnewline \cline{4-7}
 & gram  & pad  & Best & Average & Best & Average\tabularnewline
\hline \hline SMS & With & -- & 1.90 & 2.03 $\pm$ 0.05 & 1.54 & 1.64
$\pm$ 0.03\tabularnewline \cline{2-7}
 & W/O & 2.41 & 2.40 & 2.50 $\pm$ 0.04 & 1.98 & 2.02 $\pm$ 0.01\tabularnewline
\hline FCB & With & -- & 2.07 & 2.14 $\pm$ 0.03 & 1.71 & 1.78 $\pm$
0.02\tabularnewline \cline{2-7}
 & W/O & 2.37 & 2.40 & 2.54 $\pm$ 0.04 & 2.01 & 2.04 $\pm$ 0.01\tabularnewline
\hline ART & With & -- & 2.12 & 2.24 $\pm$ 0.04 & 1.77 & 1.82 $\pm$
0.02\tabularnewline \cline{2-7}
 & W/O & 2.66 & 2.53 & 2.65 $\pm$ 0.05 & 2.07 & 2.10 $\pm$ 0.01\tabularnewline
\hline
\end{tabular}
}
\end{centering}
\caption{Comparison with Moradi, using Moradi Metric}
\label{table:moradi_result1}
\end{table}

We first compare our approach with Moradi and Nickabadi.
Table~\ref{table:moradi_result1} compares five layouts using
Moradi's ~\cite{moradi:genetic} cost function: ABC, random
(with/without multigrams), and optimized (with/without multigrams),
on the three archives in Table~\ref{table:archive}. Optimized
layouts with multigrams always perform best, both in best found and
average. The ABC layout has similar (slightly worse) efficiency than
random layouts. Optimized layouts show better efficiency in every
setting, just as Moradi suggests. Efficiency is further improved by
multigrams, in both random and optimized layouts.
These results are generally consistent with those of
Moradi and
Nickabadi~\cite{moradi:genetic} (standard layout  2.90,
optimized layout 2.25), based on part of
"Harry Potter 5 - The order of the Phoenix."
The exact value
depends on the text, but in general,  optimization
based on Moradi's metric is effective in reducing typing costs for a
mobile phone keypad.

Moradi also mentions that 1.70 is the theoretical optimal value.
However this theoretical value is based on a single character per
position/strokes combination. In Table~\ref{table:moradi_result1},
we can see that multigram keypads can approach and even exceed this
theoretical optimum (in SMS). A restriction to single-character
prevents $f_1$ and$f_4$ being less than 1. With multigrams, our
design allows up to 3 characters per stroke, so $f_1$ can be below
1. $f_4$ can also be less than 1, because multigram use can
eliminate thumb movements required to type the characters within the
multigram. For example, assume that we type the word "this." Without
multigrams, we have to pay at least 1 point for each stroke. (If
some characters are on the same key, the delay penalty is 3, larger
than a 1 point move.) So, the minimum cost of thumb movement is 4.
With multigrams "th" and "is," however, we need only two movements,
so the minimum thumb movement cost is 2. Thus $f_4$ for this string
is 0.5 (movement cost 2 divided by length of the string 4). In sum,
our tests show that the multigram design can exceed the theoretical
optimal value of 1.70 suggested by Moradi.

To statistically confirm the effectiveness of multigrams for
optimizing a keypad, we provide a comparison using the Wilcoxon rank
sum test (Mann-Whitney U test), with $m = n = 50$. By the central
limit theorem, the sample approximates a normal distribution.
Table~\ref{table:wilcoxen} summarizes the test, listing sample size
$m$ and $n$, rank sum for without-multigram case $W$, its mean
$E(W)$ and standard deviation $\sigma(W)$, its normalized value
$Z_w$, and its significance level $P(Z > Z_w )$. Each column tests
the hypothesis that use of multigrams makes no difference to the
efficiency of that keypad. \footnote{For Moradi's fitness function,
all samples from keypads with multigrams have lower cost than those
without, so $W$ has the maximum possible value, 3775.} From the
normalized $Z_w$ values, the probabilities that these results arise
are all less than $10^{-15}$.

\begin{table}
\begin{centering}
{\footnotesize
\begin{tabular}{|c|r|r|r|r|r|r|}
\hline
 & \multicolumn{3}{c|}{Moradi's} & \multicolumn{3}{c|}{Two-thumb}\tabularnewline
 & \multicolumn{3}{c|}{Cost Function} & \multicolumn{3}{c|}{Cost Function}\tabularnewline \cline{2-7}
    & SMS
    & FCB
    & ART
    & SMS
    & FCB
    & ART \tabularnewline \hline \hline
 $W$ & 3775 & 3775 & 3775 & 3761 & 3755 & 3726\tabularnewline \hline
 $E(W)$ & 2525 & 2525 & 2525 & 2525 & 2525 & 2525\tabularnewline \hline
 $\sigma(W)$ & 145.1 & 145.1 & 145.1 & 145.1 & 145.1 & 145.1\tabularnewline \hline
 $Z_w$ & 8.62 & 8.62 & 8.62 & 8.52 & 8.48 & 8.28\tabularnewline \hline
 $P$ & \multicolumn{3}{c|}{$< 10^{-15}$} & \multicolumn{3}{c|}{$<
 10^{-15}$} \tabularnewline
 \hline
\end{tabular}
}
\par\end{centering}
\caption{Wilcoxon Rank Sum Test on Multigrams \protect\\
{\footnotesize $m=50$, $n=50$. $W$ is rank sum for no-multigram
case. $E(W)$ and $\sigma(W)$ are mean and standard deviation of W.
$Z_w$ is significance level, and $P$ value shows probability that
$P(Z > Z_w )$.}} \label{table:wilcoxen}
\end{table}

\subsubsection{Optimization for Two-Thumb Typing}

Moradi's function assumes we use only one thumb, so the average
movement distance is important. With two thumbs, minimizing
consecutive use of the same hand is more important. We also use
different weight factors. We can see from
Table~\ref{table:my_result1} that the current ABC keypad has much
the lowest efficiency -- far worse than random layouts. The
optimization algorithm demonstrates its effectiveness, through the
improvement from random to optimized layouts. Multigrams further
improve the efficiency, for all texts.
Again, all results are significant below the $10^{-15}$ level.

\begin{table}[bp]
\begin{centering}
{\footnotesize
\begin{tabular}{|c|l|c|c|c|c|c|}
\hline Text & Multi- & ABC & \multicolumn{2}{c|}{Random Keypad} &
\multicolumn{2}{c|}{Optimized Keypad}\tabularnewline \cline{4-7}
 & gram & pad & Best & Average & Best & Average\tabularnewline
\hline \hline SMS & With & -- & 1.41 & 1.57 $\pm$ 0.05 & 1.06 & 1.19
$\pm$ 0.04\tabularnewline \cline{2-7}
 & W/O & 1.90 & 1.65 & 1.90 $\pm$ 0.07 & 1.34 & 1.35 $\pm$ 0.01\tabularnewline
\hline FCB & With & -- & 1.54 & 1.63 $\pm$ 0.03 & 1.23 & 1.27 $\pm$
0.01\tabularnewline \cline{2-7}
 & W/O & 1.86 & 1.78 & 1.94 $\pm$ 0.06 & 1.35 & 1.36 $\pm$ 0.01\tabularnewline
\hline ART & With & -- & 1.58 & 1.71 $\pm$ 0.04 & 1.19 & 1.24 $\pm$
0.01\tabularnewline \cline{2-7}
 & W/O & 2.01 & 1.77 & 1.94 $\pm$ 0.08 & 1.30 & 1.30 $\pm$ 0.01\tabularnewline
\hline
\end{tabular}
}
\par\end{centering}
\caption{Comparison Using Two-Thumb Metric} \label{table:my_result1}
\end{table}

\subsubsection{Influence of Text Archive}
For both fitness functions, the algorithm found better fitness
values for SMS than for FCB or ART. In most cases, SMS is
significantly better than FCB, and FCB is slightly better than ART.
Thus the algorithm is especially relevant to text such as SMS on
mobile phones -- the form of text that dominates use of these
keypads. Specifically, the hypothesis that use of words and patterns
in language are more highly personalized in SMS appears correct.
Improvement is also seen in FCB and ART, but the magnitude is lower.
Gong and Tarasewicz~\cite{gong:alphabetical} also make such a
comparison, but their design shows lower efficiency on SMS than on
written or spoken language. In this respect, our algorithm is far
more effective for SMS on mobile phones.

\subsubsection {Detailed Analysis of Keypad Examples}

Table~\ref{table:detail} details our best layout for each archive.
Introducing multigrams improves every aspect affecting fitness. The
average number of strokes is reduced by 16.5\% (SMS), 7.4\% (FCB),
and 6.7\% (ART). For SMS, we average 1.01 strokes per character
 -- approximately one
character per stroke, despite the ambiguous layout.
Gong~\cite{gong:alphabetical} suggests the minimum number of strokes
per character is 1.05 for written and spoken language and 1.25 for
SMS, for 10 keys. Our layout substantially improves this -- 1.01 for
SMS and 1.12 for written language (ART).

\begin{table}[htbp]
\begin{centering}
{\footnotesize
\begin{tabular}{|c|l|c|c|c|c|}
\hline
Text & Multi- & Overall & Mean & Same
& Same\tabularnewline
 & gram & Fitness & Strokes & Key & Hand \tabularnewline
\hline \hline
SMS & With & 1.06264 & 1.01880 & 0.00614 & 0.13846\tabularnewline
\cline{2-6} & W/O & 1.34273 & 1.21908 & 0.05531 & 0.16269\tabularnewline
\hline FCB & With & 1.23325 & 1.13813 & 0.03661 & 0.16076\tabularnewline
\cline{2-6} & W/O & 1.35152 & 1.22774 & 0.05567 & 0.16106\tabularnewline
\hline ART & With & 1.19618 & 1.12686 & 0.01751 & 0.17214\tabularnewline
\cline{2-6} & W/O & 1.30019 & 1.20973 & 0.03118 & 0.17471\tabularnewline
\hline
\end{tabular}
}
\par\end{centering}
\caption{Fitness Details of Best Solutions}
\label{table:detail}
\end{table}

\begin{figure}
\centering{}
\includegraphics[width=8.5cm]{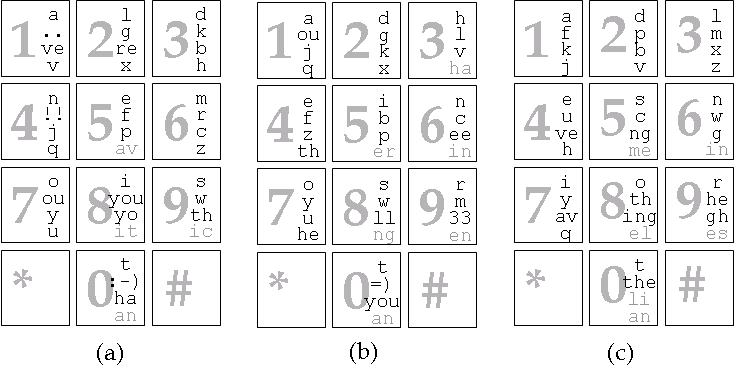}
\caption{Optimized Keypad Layouts from (a) SMS (b) FCB (c) ART}
\label{fig:keypad_example}
\end{figure}

Figure~\ref{fig:keypad_example} shows the corresponding
layouts.\footnote{Light-colored multigrams are deprecated, as they
do not improve typing speed compared to use of single characters.}
In SMS, "u" is
located on a 4-stroke position, although it is a
frequent letter in English, because the layout covers most
uses with multigrams ("ou" and "you").
In FCB, six multigrams are deprecated, but non-letters on the layout
significantly improve efficiency. For example, "33" may be used for
common emoticons like "$<:$33333." This layout also contains another
emoticon "=)". Special characters and case shifts generally require
mode changes, so directly supporting these phrases in the layout
significantly improves efficiency. In ART, we find "the" and "ing"
-- the commonest trigrams in general English -- because ART reflects
general characteristics of English.

\section{USER STUDY}
\label{sec:human}

Although this personalized-multigram (PM) layout seems
efficient, we need to verify its speed in real use, and
test its acceptability in ease of use, learnability,
and memorability, compared with the ABC-layout.

\subsection{Evaluation Design}

\subsubsection{Procedure}
\begin{figure}[tbp]
\centering{}
\includegraphics[height=5cm]{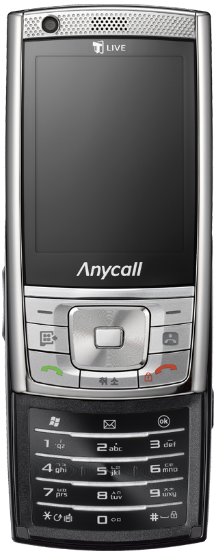}
\includegraphics[height=5cm]{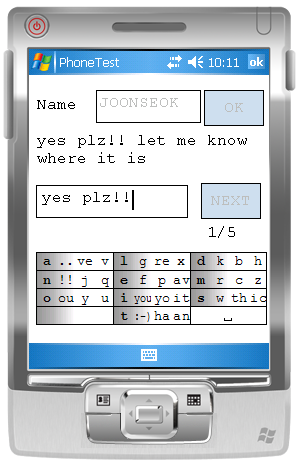}
\caption{ The phone (left) and interface (right) used in human
evaluation. } \label{fig:phone}
\end{figure}

The overall idea is to compare typing speed between PM and ABC by
measuring elapsed time for each participant typing sample messages
displayed on the screen. We used the same device (Figure~\ref{fig:phone}
left) and same program UI (Figure~\ref{fig:phone} right) for testing
both layouts. Only the keypad layout shown on the bottom was different.
The programs for the two layouts were compiled separately, yielding two different
executables.

To observe the after-training speed and verify learnability, we
repeat the experiment for several sessions. A session consists of a
set of 5 English text messages of 15 to 30 bytes. Each participant
is instructed to repeat at least 10 times, but they are free to do more until
they feel no further improvement is possible for each layout. In
order to simulate casual use of text messages, the experiment is
spread over 2 to 3 days for each participant. Participants are guided to
try any time they wish during the test period. After the completion
of the sessions, a questionnaire is distributed to check their
satisfaction level and subjective feeling of improvement. Messages
are randomly selected from a pool of real SMS texts, because this
experiment aims to evaluate a mobile phone layout. A few examples:

$\circ$ have a nice day :-)\\
$\circ$ omg.. take more than a month..\\
$\circ$ heyhey i got a msg from her!!

The program interface is shown in Figure~\ref{fig:phone}. After
entering name and pressing the START button, it displays a
message to type. For each message, it records typed text and
time in milliseconds, from the first letter press
to the OK button (on the keypad). For accurate
measuring, we instructed subjects not to pause during a
message. Using the records, we measure the subject's typing speed
and accuracy as the average typing time per character, imposing a
penalty of 1 second per typo
(calculated as the
Levenshtein Distance~\cite{leven:ldist} between target
and answer).
For the unfamiliar PM layout, the key arrangement is shown on the screen. \footnote{There is
some cognitive load in this,
imposing a slight bias against the proposed layout; in light
of the results, this bias was not serious.}

\subsubsection{Subjects}
We tested 10 subjects, 9 male and 1 female. Most were researchers in
computer science except for two. Seven were in their mid-20s, the
other three being 33, 53, and 61. Average age was 32.8. Eight were
Korean, the other two being from Australia and Spain.

This experiment required multiple trials over about 3 days for each
participant, so it took a lot of time. In order to reduce the
elapsed time, we shared the phone between people who used the same
work space all day. A university laboratory was chosen for this
reason, although it led to some bias in gender and occupation.
However, we strictly controlled experience and familiarity with the
current mobile phone keypad, the most important factor for this
experiment.

Note that our participants include two seniors, who showed
dramatically slower improvement. For homogeneity of results, we
excluded them from the analysis, as we show raw data in
Table~\ref{table:human_individual}. It seems clear that age is a
confounding variable that needs to be explored in more detail in
future.

\subsubsection{Apparatus}
We used a Samsung Electronics SCH-M470 (Figure~\ref{fig:phone})
because of its combination of a 12-key standard alpha-numeric
physical keypad with programmable smart-phone capabilities, making
easy to install and run our test program. (Most other smart-phones
use a touch-screen input method, which is inappropriate for this
experiment.) The operating system is Microsoft Windows Mobile
Version 6.0 Professional. The phone size is $101.5 \times 53 \times
16.8$ mm.
The program was developed in C++, using Microsoft Visual Studio
2008.

\subsection{Results and Analysis}

\begin{table}[tbp]
\begin{centering}
{\footnotesize
\begin{tabular}{|l|r|r|r|r|r|r|r|}
\hline No. & \multicolumn{3}{c|}{Typing Speed (CPM)} & Speed &
Cross- & Session & Age \tabularnewline \cline{2-4}
 & ABC & PM & PM & Up & point & Count & \tabularnewline
&  & First & Final & & & & \tabularnewline \hline \hline 
1 & 36.47 & 22.60 & 76.87 & 111\% & 3 & 14 & 26\tabularnewline \hline 
2 & 35.01 & 28.65 & 74.65 & 113\% & 3 & 17 & 33\tabularnewline \hline 
3 & 55.20 & 49.42 & 104.50 & 89\% & 4 & 22 & 27 \tabularnewline \hline 
4 & 65.21 & 29.64 & 82.91 & 27\% & 3 & 12 & 26\tabularnewline \hline 
5 & 30.43 & 20.40 & 37.79 & 31\% & 3 & 7 & 28\tabularnewline \hline 
6 & 72.84 & 50.09 & 100.30 & 38\% & 7 & 23 & 25\tabularnewline \hline
7 & 37.26 & 22.66 & 46.43 & 25\% & 6 & 9 & 24\tabularnewline \hline
8 & 36.52 & 14.28 & 45.13 & 24\% & 19 & 34 & 61\tabularnewline \hline 
9 & 33.37 & 19.13 & 50.26 & 51\% & 17 & 22 & 53\tabularnewline \hline 
10 & 46.49 & 32.40 & 71.73 & 54\% & 4 & 20 & 25\tabularnewline \hline\hline 
Avg & 44.88 & 28.93 & 69.26 & 54\% & 6.90 & 18.00 & 32.8\tabularnewline \hline
SD & 14.69 & 12.22 & 23.43 & 35\% & 6.03 & 7.97 & 13.1\tabularnewline \hline
\end{tabular}
}
\par\end{centering}
\caption{Individual results in human evaluation\protect\\
{\scriptsize Typing speeds are in CPM (Character per minute), and
PM means Personalized-Multigram. Speed Up denotes the percent
speedup between PM-layout and ABC-layout typing speed. Cross-point
denotes the first PM session at which the subject exceeded their
ABC-layout speed.}} \label{table:human_individual}
\end{table}

\subsubsection{Result}

\begin{figure}[htbp]
\centering{}
\includegraphics[width=8.5cm]{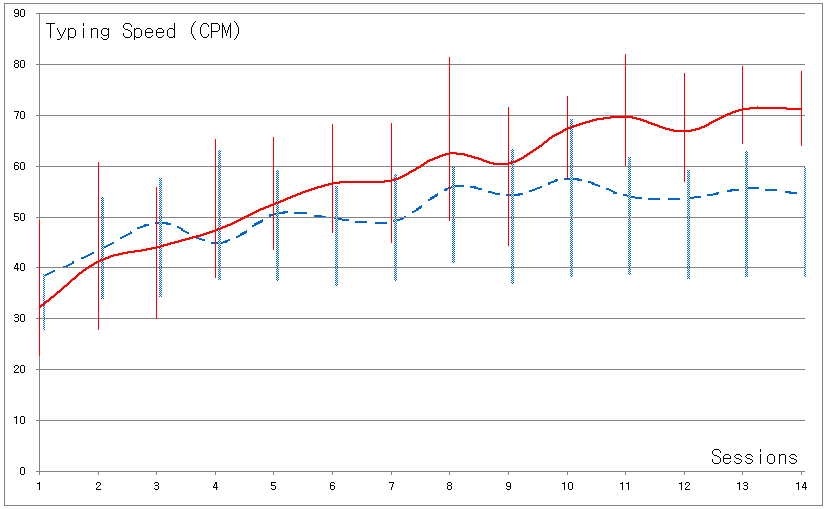}
\caption{ Overall progress for both layouts across sessions.
Red line is for PM, blue-dotted line for ABC. The
vertical lines show the range of speeds. The
progress line connects the averages. }
\label{fig:result}
\end{figure}

Figure~\ref{fig:result} shows an increasing trend in typing speed for
both layouts. Each point is the average of all subjects' typing speed
at that session. At first, ABC speed (38.39 CPM)
exceeds PM (32.22) --
probably due to previous familiarity. By session
4, PM (47.41) speed
overtakes ABC (44.72). Eventually, PM reaches 71.21
while ABC stagnates around 54. Over 14 sessions, subjects showed
121\% improvement with PM, but only 42\% with ABC.
Individually, this ranged from 86\% to 240\% for PM,  from
5\% to 59\% for ABC.
Subjects show better
performance with PM after only a few sessions, confirming PM's
efficiency.

In Table~\ref{table:my_result1}, we saw that the typing cost for SMS
reduced from 1.90 (current layout) to 1.06 (optimized Multigram
layout) -- around a 44.2\% improvement. Results from the human
evaluation suggest that this is in the right ballpark. Specifically,
subjects in the experiment averaged around 52.9\% reduction in
typing time, ranging from 14.3\% to 80.8\%.

\subsubsection{Usability Issues}

Usability is an important issue. A user familiar with ABC layout may
not be prepared to make the effort required for another; a new user
may prefer a layout that appears familiar (like ABC). In order to
gauge this effect, we distributed a questionnaire to each subject
after completing all sessions. Each question used a six-point Likert
Scale format: marking any point between 0 and 5.

We investigated usability by asking users' personal feelings about
both personalized layout and multigrams. We asked which is
preferable between the personalized (5 point) vs. standardized
layout (0 point). Average response was 3.25: the personalized pad
was slightly preferable, but similar. For multigrams users showed
greater satisfaction. Between perfectly satisfied (5 point) and not
satisfied (0 point), the average response was 4.05. The final
question asked users to choose between two options,
personalized-multigram layout and current ABC-layout, assuming that
no other layouts were available. We gave three options: PM-layout,
ABC-layout, and 'do not matter'. Six subjects out of 10 chose
PM-layout, two ABC-layout and two 'does not matter'. To sum up, we
can conclude that subjects tended toward satisfaction with the
personalized-multigram layout, in spite of its complete
unfamiliarity.

Usability can be metricized for both
learnability and memorability. We infer that the PM-layout is fairly
easy to learn by investigating the overtaking point (first time that PM
shows better performance). On average, each subject exceeded their own
ABC speed in 4.7 sessions, ranging from 3 to 7. Early sessions
take about 5 minutes, so they were able to learn PM
layout sufficiently in about 30 minutes. We also asked about this in
the questionnaire: "do you think you have improved in remembering
each key's location in the new layout, compared to the early stages
of the experiment?" (very much - 5 point to not improved at all - 0
point.) For this question, almost everyone responded that they have
improved, indicated by the average of 4.00. We conclude that PM layout
is sufficiently easy to learn.

\section{DISCUSSION}
\label{sec:disc}
\subsection{Contributions}
The personalized multigram layout promotes much higher typing speed
on mobile phones through two key features: an efficient arrangement
of letters reflecting individual needs, and introduction of
frequently-used multigrams. If the system were offered by
manufacturers as a table-driven alternative, then users could
generate the layout from a sample of their text, and upload it to
any phone offering the feature. Of course, users would always have
the option to use a table giving them the current (ABC) layout.

Some have voiced concern that phone keypads are being replaced by
touch screens, vitiating this work. 
This may not reflect a truly international mindset.
It is true that smart-phone use in the first
world is rapidly increasing. According to market researcher
In-Stat~\cite{trend:smartphone}, the number of smart-phone users
world-wide in 2009 was estimated at about 25 million, expected to
increase four-fold by 2013. However, the total number of mobile
phone users was about 3.1 billion in 2007, expected to increase to
4.5 billion in 2012.~\cite{trend:mobilephone}. Thus keypad-based
phones are still increasing in number, and will continue to increase
over the immediate future; a tailorable keypad layout will continue
to have value for a substantial time, and could offer market
advantage to manufacturers. Equally important, the general
philosophy of personalized layouts, optimized to a specific user's
pattern of use, is even more readily applicable to touch-screen
smartphones than to keypad phones: the key entry is already purely
software-based, so a tailorable system is easy to deploy.

\subsection{Limitations}
Although the personalized multigram layout has merit, it suffers
from some limitations. Firstly, for a fixed layout, it has been the
practice to engrave the letters on their corresponding keys. For a
personalized layout, this is not so easy. It will generally be
necessary to display the layout on the screen, at least while the
user is learning it. This may restrict the use of screen space for
applications. In typical phones, perhaps one third of the screen
would be occupied by the layout, leaving only two thirds for
applications. It might be possible to reduce this disadvantage
through stick-on labels or other means, but it needs to be
acknowledged.

Second, there is a small possibility that some hardware might need
re-design to support this function, because the layout would need to
be stored in memory. We suspect that current keypad phones actually
do this anyway, to support easy customization to different alphabets
in different markets. But this design information is not readily
available from manufacturers, so we have been unable to confirm it.

Third, language usage patterns may change over time. In our proposed
use model, the layout is fixed once it has been created, so it may
become out-of-date, not reflecting the user's pattern as well as it
once did. Of course it is possible for users to re-run the algorithm
at any time with a new sample of their text, and generate a new
layout. However the new layout would impose a substantial challenge
-- re-learning.  We address this further below.

Fourth, because of the high effort involved in testing, the user study 
we conducted, like all such studies in the literature, used a very small 
sample size (10) -- too small for reliable statistical testing. However
even ignoring the very small sample size, and treating the data as
Gaussian, yields a confidence interval at the $5\%$ level that includes
zero: even at a $5\%$ confidence level, and making very optimistic
assumptions, we cannot exclude the possibility that there
is no difference between PM and ABC layout speed. We suspect that
this study is not alone in this respect, and that few such learning-based
studies in the literature could withstand a detailed significance 
analysis.

Last (and least important), it takes some time to compute a
new  layout. In our computer system,
it took about 3 to 5 minutes. If run directly on a mobile phone, it
would take more time. Even in this case, it is required only
infrequently, and could be run overnight during sleep.

\subsection{Further Work}
Changing patterns of use may change the optimal  layout. But
complete re-optimization is likely to generate very different
layouts, imposing high learning loads on the user. Alternatively,
the amount of change from the previous layout could be incorporated
in the fitness function, to minimize the cognitive load while
maintaining high typing speed. In general, the issue of how to
maintain high typing speed in the face of changing use patterns,
while also minimizing cognitive load, is a challenging research
topic (not only for mobile phone keypads), worthy of a substantial
research investment.

We would like to extend our experiments to a larger number of people
with more varied backgrounds. Our current testing is somewhat
limited in the sense of statistical stability. A large proportion of
our subjects are young, male CS-majored graduate students. It
appears likely, both from experience and from our experiments, that
age is a strong determinant of learning. We need to test with more
people with a wider range of characteristics, but the current
protocol imposes a substantial effort on subjects, and over a long
time. To overcome this, we have designed a new experimental
protocol. We will be running detailed human experiments with a
larger and more varied pool of subjects in the near future.

\section{CONCLUSION}
\label{sec:conc}

In this paper, we described an optimization algorithm for
personalized mobile phone key layout using a multigram approach.
Genetic algorithms are used to find both the best combination of
multigrams, and the best key arrangement. The fitness function takes
into account the average number of strokes required to type text,
consecutive strokes on the same key, and consecutive use of the same
hand. Applying this algorithm, we have found highly efficient
multigram layouts for three kinds of archives: instant messages as
used in cell phones; those used in social network services; and
general articles. Fitness values for these best layouts are
significantly smaller than for layouts without multigrams,  and for
unoptimized layouts -- including the current alphabetically-ordered
layout.

The general method is almost independent of language or character
set -- it can be applied to any single-level alphabetic language.
With minor adaptations, it can handle multi-level alphabetic
languages such as Korean, and unambiguous representations of
ideographic languages (e.g. wu-bi method for Chinese), but not
ambiguous representations (pinyin).

In comparison with previous work, we show that our design
substantially improves efficiency. The advantage is particularly
great for SMS texts. Human evaluation verifies its usability and
learnability, and its impact on performance.

%#################################### Begin Main Content ##############################################
 
\bibliographystyle{abbrv}
\bibliography{joonseok,bob}

\end{document}